# Investigation of superconducting properties and possible nematic superconductivity in self-doped BiCh$_2$-based superconductor CeOBiS$_{1.7}$Se$_{0.3}$

*Ryosuke Kiyama*[1], *Kazuhisa Hoshi*[1], *Yosuke Goto*[1], *Yoshikazu Mizuguchi*[1]*

1. Department of physics, Tokyo Metropolitan University, 1-1, Minami-osawa, Hachioji, 192-0397, Japan.

E-mail: mizugu@tmu.ac.jp

**Abstract**

We investigate the superconducting properties and possible nematic superconductivity of self-doped BiCh$_2$-based (Ch: S, Se) superconductor CeOBiS$_{1.7}$Se$_{0.3}$ through the measurements of in-plane anisotropy of magnetoresistance. Single crystals of CeOBiS$_{1.7}$Se$_{0.3}$ were grown using a flux method. Single-crystal structural analysis revealed that the crystal structure at room temperature is tetragonal (P4/*nmm*). Bulk superconductivity with a transition temperature of 3.3 K was observed through electrical resistivity and magnetization measurements. Investigation of anisotropy of upper critical field suggested relatively low anisotropy in the crystal as compared to other BiCh$_2$-based superconductors. In the superconducting states of CeOBiS$_{1.7}$Se$_{0.3}$, two-fold symmetric in-plane anisotropy of magnetoresistance was observed, which indicates the in-plane rotational symmetry breaking in the tetragonal structure and hence the possibility of nematic superconductivity in CeOBiS$_{1.7}$Se$_{0.3}$.

# 1. Introduction

Recently, nematic superconductivity (NSC), which is typically characterized by spontaneous rotational symmetry breaking in the amplitude of the superconducting gap, has been observed in superconducting states of a topological superconductor system $A_x$Bi$_2$Se$_3$ (A = Cu, Sr, Nb).[1-7] Although $A_x$Bi$_2$Se$_3$ has a trigonal structure, the observed in-plane anisotropy of superconducting properties in those superconductors exhibit two-fold symmetry. This behavior indicates rotational symmetry breaking in the superconducting states and is called NSC. Since the system which shows NSC is limited, new system which shows NSC has been desired to obtain further knowledge about the emergence of NSC states in layered materials.

Very recently, in single crystals of LaO$_{1-x}$F$_x$BiSSe ($x$ = 0.1 and 0.5), which are BiCh$_2$-based layered superconductors (Ch: S, Se),[8-10] two-fold symmetric in-plane anisotropy of c-axis magnetoresistance (MR) was observed in the superconducting states,[11,12] while those phases have tetragonal symmetry.[13] Those experimental results proposed BiCh$_2$-based superconductors to be a new platform to study NSC. Since both samples with different electron doping concentration ($x$ = 0.1 and 0.5) exhibit NSC, Fermi surface topology seems not essentially affect the emergence/disappearance of NSC in LaO$_{1-x}$F$_x$BiSSe. However, in another BiCh$_2$-based superconductor, NdO$_{0.7}$F$_{0.3}$BiS$_2$, no features of NSC were observed in its superconducting states; MR in the superconducting states exhibited four-fold-symmetric in-plane anisotropy.[14] Therefore, we need to clarify the factor, which is essential for the emergence/disappearance (switching) of NSC in BiCh$_2$-based systems. For that purpose, comparison of properties of phases with different in-plane chemical pressure (CP) should be important.[15] Generally, superconducting properties of BiCh$_2$-based systems are explained by in-plane CP because their superconducting properties are strongly affected by the in-plane local disorder, which locally lowers in-plane structural symmetry of the Bi-Ch1 network (see **Figure 1(b)** for the definition of the Ch1 site).[15-20] With this respect, LaO$_{1-x}$F$_x$BiSSe possesses quite high in-plane CP because of nearly 100% occupancy of Se at the in-plane Ch1 site. For Nd(O,F)BiS$_2$, in-plane CP is lower than that for LaO$_{1-x}$F$_x$BiSSe,[15] while it is higher than other RE(O,F)BiS$_2$ with RE = La, Ce, and Pr and hence shows bulk superconductivity. Therefore, there is possible scenario that the presence of weak local structural disorder suppresses the NSC states in BiCh$_2$-based systems. Hence, we need further example of BiCh$_2$-based superconductors, which shows NSC for clarifying the scenario above.

This study is focused on the CeOBiS$_{2-x}$Se$_x$ system. Bulk superconductivity was observed in polycrystalline samples with $x$ = 0.4 and 0.6 of CeOBiS$_{2-x}$Se$_x$, and its $T_c$ was ~3 K.[21] The crystal structure is composed of alternate stacks of a CeO blocking layer and a BiCh$_2$ conducting layer. Notably, external elemental substitution is not needed to generate electron carriers in BiCh$_2$ layers because carriers are self-doped via the mixed valence states of Ce.[21,22] This situation is clearly different from the cases of other RE(O,F)BiCh$_2$ systems, in which elemental substitutions are needed to dope electrons.[9,10] In addition, Se concentration in Bi-Ch1 conducting plane is significantly lower than LaO$_{1-x}$F$_x$BiSSe. Hence, there are clear compositional differences in both blocking and conducting layers between LaO$_{1-x}$F$_x$BiSSe ($x$ = 0.1 and 0.5), NdO$_{0.7}$F$_{0.3}$BiS$_2$, and CeOBiS$_{2-x}$Se$_x$. In this study, we have investigated the in-plane anisotropy of MR in the superconducting states of CeOBiS$_{2-x}$Se$_x$ ($x$ = 0.3) and observed NSC features.

## 2. Experimental details

CeOBiS$_{2-x}$Se$_x$ single crystals were grown by using a high-temperature flux method in an evacuated quartz tube. Polycrystalline powders of CeOBiS$_{1.6}$Se$_{0.4}$ were prepared by the solid-state-reaction method as described in Ref. 21. The mixture of the polycrystalline powders of CeOBiS$_{1.6}$Se$_{0.4}$ (~0.5 g) and CsCl flux (~3.8 g) with a molar ratio of 1:20 was mixed and then sealed into an evacuated quartz tube. The tube was heated at 950 °C for 15 h and slowly cooled to 650 °C at a rate of −2.0 °C/h, which was followed by furnace cooling to room temperature. After the heat treatment, the quartz tube was opened under the air atmosphere, and the product was filtered and washed with pure water.

The obtained samples were characterized by X-ray diffraction method (XRD) with a Cu-K$_\alpha$ radiation on a Miniflex-600 (Rigaku) by the $\theta$–$2\theta$ method. The single crystals were analyzed by scanning electron microscopy (SEM), and their chemical composition was investigated by energy-dispersive X-ray spectroscopy (EDX). Single crystal X-ray structural analysis was carried out at room temperature on a XtaLAB (Rigaku). The structural parameters were refined with the tetragonal (P4/*nmm*) structural model using the refinement program SHELXL.[23] A crystal structure image was depicted using a VESTA software.[24]

Electrical resistivity was measured by an in-plane four-terminal configuration. The terminals were fabricated using Au wire (25 µm) and Ag paste. In-plane anisotropy was investigated on a PPMS with a horizontal rotator (Quantum Design) and a sample holder for in-plane anisotropy measurements. The temperature dependence of magnetic susceptibility was measured by a superconducting interface device (SQUID) magnetometer with an applied field 10 Oe after zero-field cooling (ZFC) and field cooling (FC) on a MPMS3 (Quantum Design).

## 3. Results and discussion

As shown in the inset of **Figure 1(a)**, plate-like crystals were obtained. To confirm the *c*-axis direction of the obtained sample, powder XRD was performed on the CeOBiS$_{1.7}$Se$_{0.3}$ single crystals; the plates of the crystal were loaded on a sample plate. As shown in Fig. 1(a), 00*l* peaks were solely observed, which confirms that the *ab*-plane is well developed. The average ratio of the constituent elements (except for O) was estimated to be Ce : Bi : S : Se = 1 : 1.00(1) : 1.74(2) : 0.30(1), where the Ce value was fixed to 1 for normalization. Considering the error in the EDX analysis, we concluded that the chemical composition of the single crystals is close to the nominal composition, and hence the sample was called CeOBiS$_{1.7}$Se$_{0.3}$ in this paper.

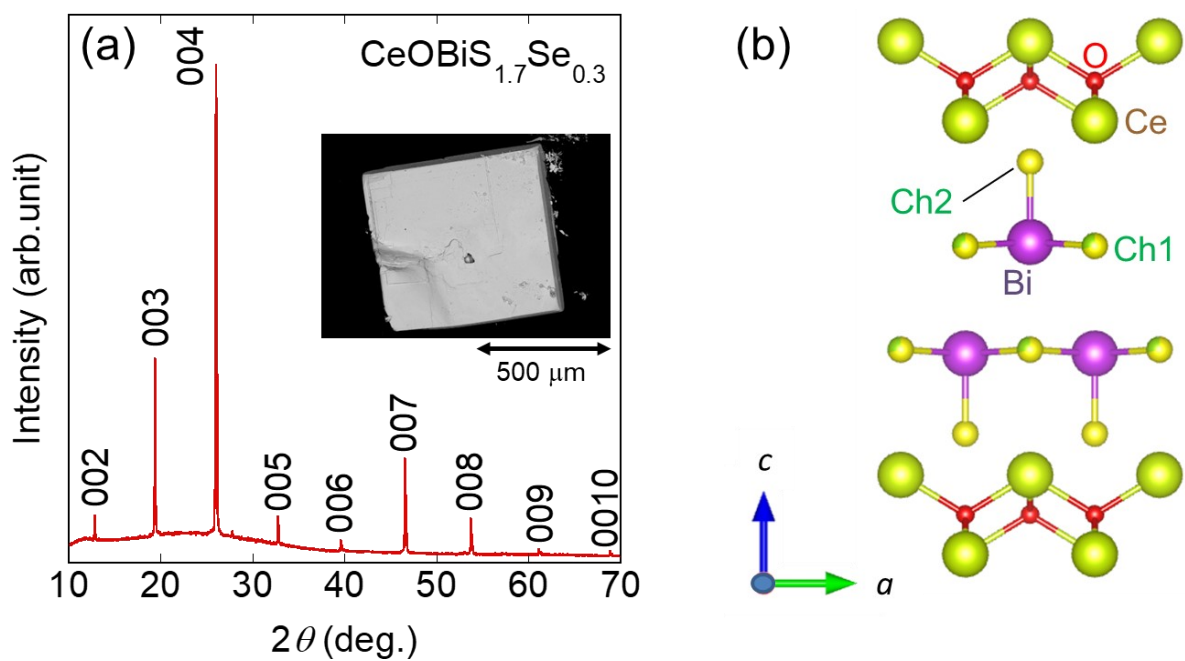

**Figure 1**: (a) XRD pattern of CeOBiS$_{1.7}$Se$_{0.3}$ single crystals. The inset shows an SEM image of a CeOBiS$_{1.7}$Se$_{0.3}$ single crystal. (b) The schematic representation of the refined crystal structure of CeOBiS$_{1.7}$Se$_{0.3}$. Ch1 and Ch2 denote the in-plane and out-of-plane chalcogen sites, respectively.

Single crystal X-ray structural analysis on a CeOBiS$_{1.7}$Se$_{0.3}$ crystal was performed. Details of the analysis condition and the refined structural parameters are summarized in **Tables I and II**. Since the Se ions selectively occupy the in-plane Ch1 site,[25] Se occupation ratio for Ch2 site fixed to 0. The ratio of S and Se is consistent with the EDX analysis result. The CeOBiS$_{1.7}$Se$_{0.3}$ single crystal crystallizes in a tetragonal space group P4/*nmm* and has lattice parameters of $a$ = 4.0327(9) Å and $c$ = 13.603(4) Å, which agrees with a report on polycrystalline samples.[21] As demonstrated in Ref. 26, the bond valence sum for Ce site was calculated using the following parameters: $b_0$ = 0.37 Å, $R_0$ = 2.151 Å for Ce–O bond, 2.62 Å for Ce–S bond, and 2.74 Å for Ce–Se bond. Bond distances between Ce and nine coordinating anions were determined by single crystal X-ray structural analysis. Site occupancies at the chalcogen site were included in the calculation. The estimated valence of Ce is 3.25, which is consistent with the value for the polycrystal and indicates that Ce has the mixed valence state.[21] Hence, the phase is a self-doped system with mixed-valence Ce.

**Figure 2** shows the temperature dependence of the electrical resistivity for a CeOBiS$_{1.7}$Se$_{0.3}$ single crystal measured with a current along the (a) *ab*-plane ($\rho_{ab}$) and (b) *c*-axis ($\rho_c$). The onset temperature ($T_c^{onset}$) and zero-resistivity temperature ($T_c^{zero}$) were determined to be 3.6 K and 3.3 K, respectively. The $T_c^{zero}$ values are slightly higher than those observed for polycrystalline samples of CeOBiS$_{2-x}$Se$_x$.[21] In the normal states, $\rho_c$ is clearly higher than $\rho_{ab}$, which is due to the structure composed of stacking of electrically conductive BiCh$_2$ layers and insulating layers along the *c*-axis. Similar behavior has been observed in various layered superconductor, such as a Cu oxide Bi$_2$Sr$_2$CuO$_x$,[27] an Fe pnictide BaFe$_2$As$_2$,[28] and BiCh$_2$-based compounds.[29]

**Table I.** Refined atomic coordinates, $B_{eq}$ and occupancy of CeOBiS$_{1.7}$Se$_{0.3}$.

| site | *x* | *y* | *z* | $B_{eq}$ (Å$^2$) | Occupancy |
|---|---|---|---|---|---|
| Bi | 0 | 0.5 | 0.62876(12) | 0.90(8) | 0.98(6) |
| Ce | 0 | 0.5 | 0.09148(19) | 0.63(9) | 1.01(6) |
| S1 | 0 | 0.5 | 0.3804(8) | 1.3(2) | 0.72(5) |
| Se1 | 0 | 0.5 | 0.3804(8) | 1.3(2) | 0.28(5) |
| S2 | 0 | 0.5 | 0.8135(8) | 0.8(3) | 1.00(9) |
| Se2 | 0 | 0.5 | 0.8135(8) | 0.8(6) | 0 (fixed) |
| O | 0 | 0 | 0 | 0.0(7) | 0.9(2) |

**Table II.** Refined structural parameters and analysis condition of $CeOBiS_{1.7}Se_{0.3}$.

| | |
|---|---|
| Formula | $CeOBiS_{1.7}Se_{0.3}$ |
| Formula weight | 443.30 |
| Space group | Tetragonal *P*4/*nmm* (#129) |
| Lattice type | Primitive |
| *Z* value | 2 |
| *a* (Å) | 4.0327(9) |
| *c* (Å) | 13.603(4) |
| *V* (Å$^3$) | 221.22(9) |
| Residuals: *R* (All reflections) | 0.0793 |
| Goodness of Fit Indicator | 1.184 |
| Crystal dimensions | 0.0830 × 0.100 × 0.010 mm |
| Diffractometer | XtaLAB mini |
| Radiation | MoK$_\alpha$ ($\lambda$ = 0.71075 Å) (graphite monochromated) |
| Temperature (K) | 293 |

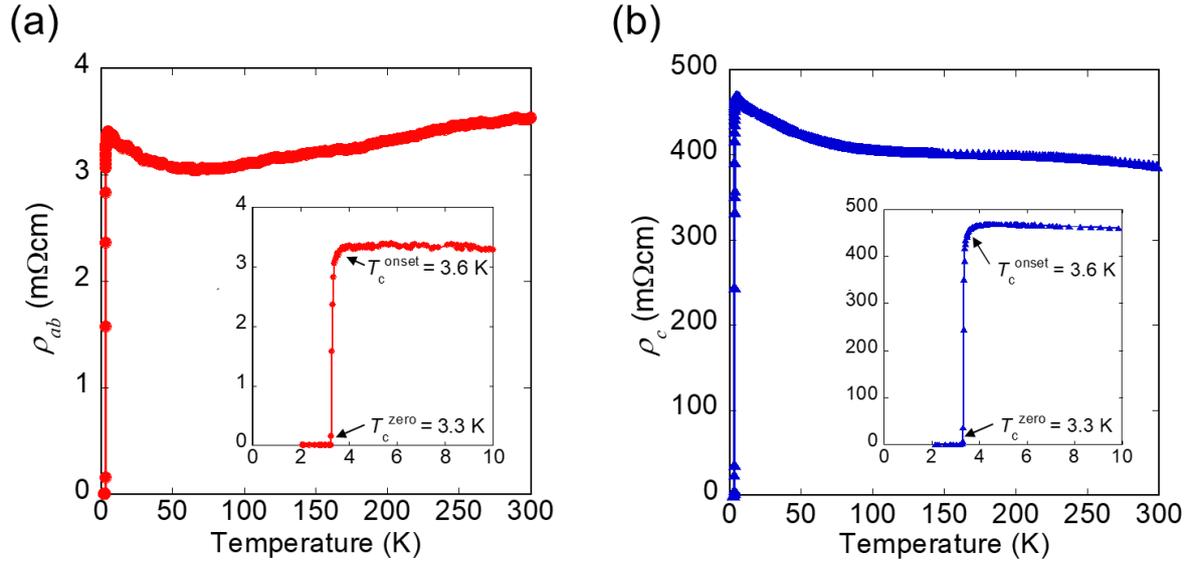

**Figure 2**: Temperature dependence of electrical resistivity for a CeOBiS$_{1.7}$Se$_{0.3}$ single crystal measured with a current along the (a) *ab*-plane ($\rho_{ab}$) and (b) *c*-axis ($\rho_c$).

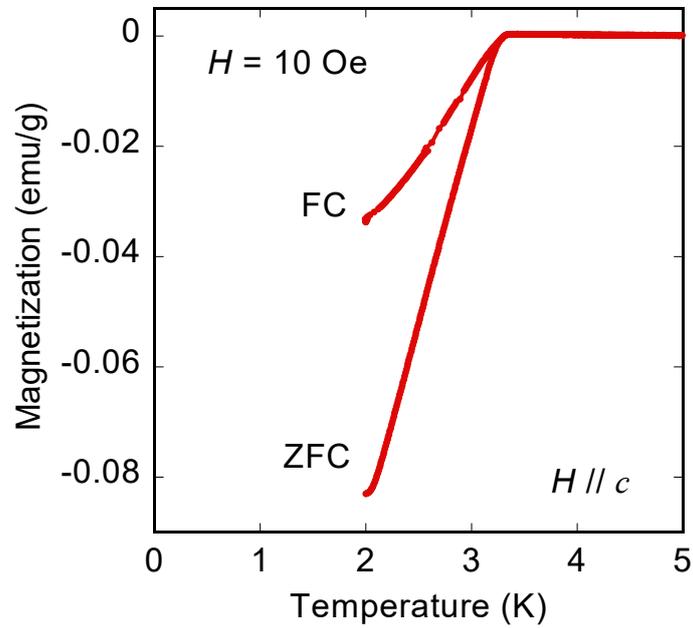

**Figure 3**: Temperature dependence of magnetization (ZFC and FC) for a CeOBiS$_{1.7}$Se$_{0.3}$ single crystal.

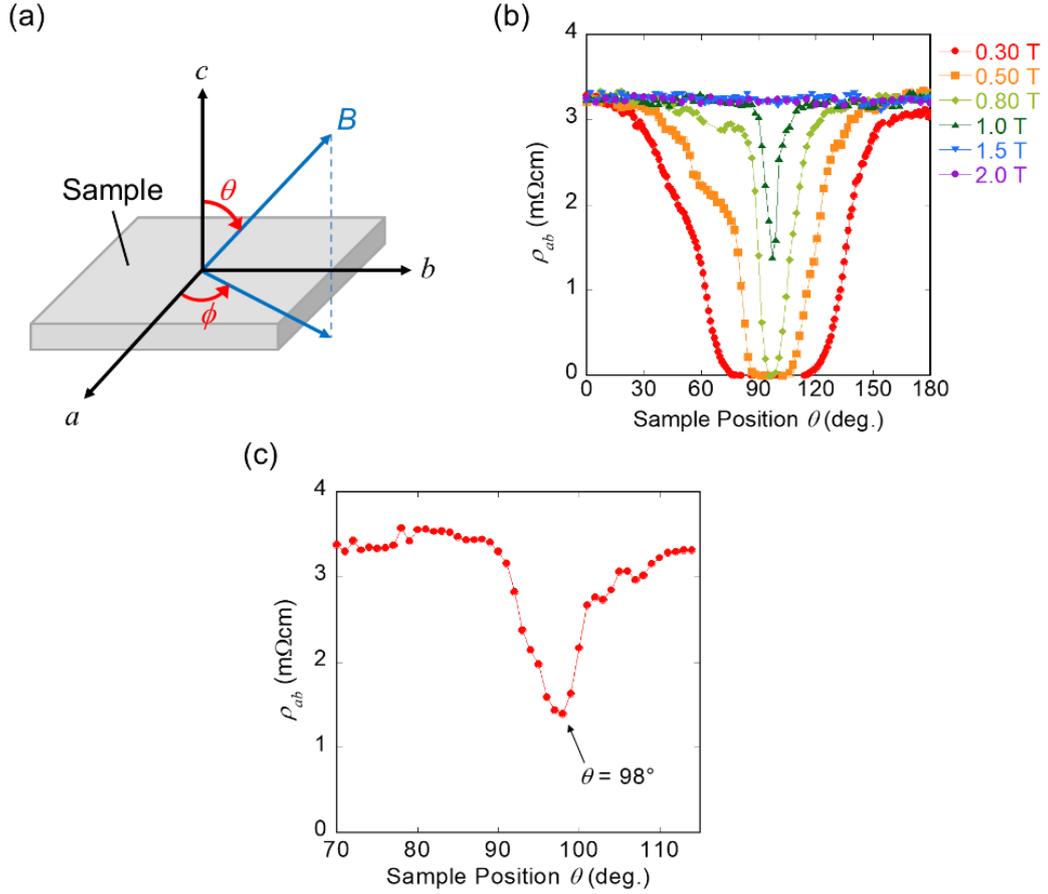

**Figure 4**: (a) Schematic image of the rotation angles. (b) $\theta$ dependence of $ab$-plane resistivity ($\rho_{ab}$) for a CeOBiS$_{1.7}$Se$_{0.3}$ single crystal under various magnetic in the range 0.30–2.0 T. (c) $\theta$ dependence of $\rho_{ab}$ for a CeOBiS$_{1.7}$Se$_{0.3}$ single crystal at $B = 1.0$ T and $T = 2.1$ K.

**Figure 3** shows the temperature dependence of the magnetization after ZFC and FC with an applied field of 10 Oe parallel to the $c$-axis for CeOBiS$_{1.7}$Se$_{0.3}$ single crystal. A large diamagnetic signal corresponding to superconductivity was observed, indicating that the observed superconducting states are bulk in nature, as reported in polycrystalline samples.[21] $T_c$ was estimated to be 3.3 K, which is consistent with the zero resistivity states in the $\rho$-$T$ data.

**Figure 4(a)** shows a schematic image of the rotation angles for magnetoresistance measurements. **Figure 4(b)** shows the $\theta$ dependence of $\rho_{ab}$ at various magnetic fields of 0.3–2.0 T. Anisotropy was clearly observed, which is typical trend for layered superconductors. **Figure 4(c)** shows the $\theta$ dependence of $\rho_{ab}$ at $B = 1.0$ T and T = 2.1 K. The minimum of $\rho_{ab}$ was observed at $\theta = 98°$, in which the magnetic field is applied parallel to the *ab*-plane. Therefore, we defined that the magnetic field are parallel to the *ab*-plane when $\theta = 98°$.

**Figure 5** shows the temperature dependence of $\rho_{ab}$ under various magnetic fields of (a) *B*//*ab* and (b) *B*//*c*. $T_c$ decreases with increasing magnetic field in both directions. The suppression of $T_c$ under magnetic field parallel to the *c*-axis is more significant than that under magnetic field parallel to the *ab*-plane. An upper critical field ($B_{c2}$)-temperature phase diagram is shown in Fig. 5(c), in which a temperature where the resistivity becomes 90% of the normal-state value under various applied magnetic fields. We calculated $B_{c2}(0)$ for *B*//*ab* and *B*//*c* using the conventional one-band Werthamer-Helfand-Hohenberg (WHH) model for type-II superconductors in a dirty limit,[30] which gives $B_{c2}(0) = -0.693T_c(dB_{c2}/dT)_{T=T_c}$. The $B_{c2}(0)$ values for *B*//*ab* and *B*//*c* were estimated to be $B_{c2}^{//ab}(0) = 3.3$ T and $B_{c2}^{//c}(0) = 0.46$ T, respectively. The anisotropic parameter for $B_{c2}$, $\gamma = B_{c2}^{//ab}(0)/ B_{c2}^{//c}(0)$, is determined to be 7.3. This value is comparably lower than that of other $BiCh_2$-based superconductors.[29] Since the anisotropy parameter depends on its fluorine concentration in the case of RE(O,F)BiS$_2$ with RE = La, Ce, Pr and Nd,[39, 31] the lack of fluorine may affect the $\gamma$ in $BiCh_2$-based systems. It is worth noting that, $\gamma$ was estimated to be 30–60 when RE = La, Pr, and Nd; the value of $\gamma$ was especially high for Pr(O,F)BiS$_2$, while it was estimated to 13–21 for Ce(O,F)BiS$_2$.[30, 31]

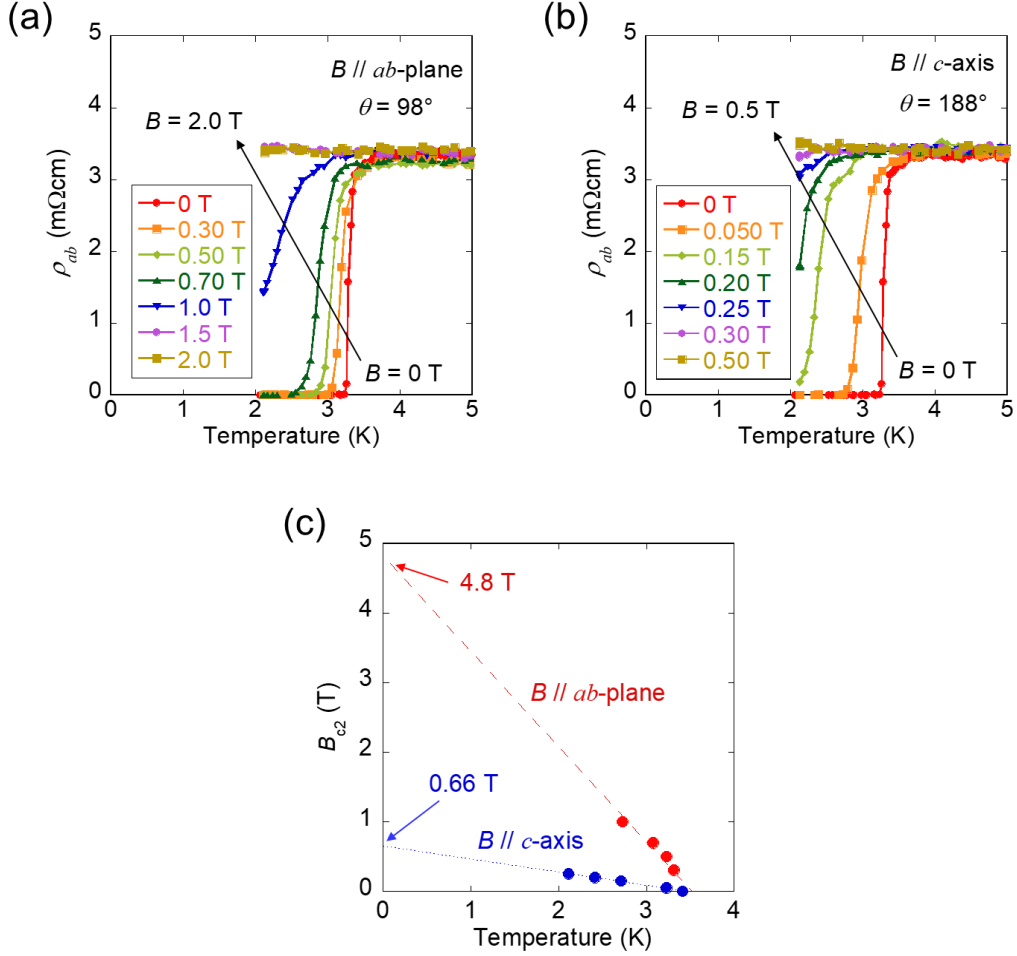

**Figure 5**: Temperature dependences of the *ab*-plane resistivity ($\rho_{ab}$) for a CeOBiS$_{1.7}$Se$_{0.3}$ single crystal under various magnetic fields along (a) *B//ab* and (b) *B//c*. (c) Temperature dependence of upper critical field $B_{c2}(T)$ for a CeOBiS$_{1.7}$Se$_{0.3}$ single crystal. The values estimated from the linear extrapolations were used for the estimation of $B_{c2}(0)$ using the WHH model.

**Figure 6(a)** shows the $\phi$ dependences of the $\rho_{ab}$ at $B = 0.5$ T and at various temperatures ranging 2.4–8.0 K. Below 3.0 K, the $\phi$ dependence of the $\rho_{ab}$ shows two-fold-symmetric in-plane anisotropy. This trend is not expected from a tetragonal structural symmetry: four-fold symmetry in a tetragonal *ab* plane. The structural-symmetry breaking in the superconducting states is a trend of NSC, which has been observed in LaO$_{1-x}$F$_x$BiSSe.[11,12] Above 4.0 K, the $\rho_{ab}$ is independent of $\phi$, which suggests that the observed two-fold symmetry of magnetoresistance appears in superconducting states only. To investigate the reproducibility and the affection of the magnitude of magnetic fields for the observed two-fold symmetric in-plane anisotropy, we performed the in-plane anisotropy measurement under various magnetic fields in a different sample. **Figure 6(b)** shows the $\phi$ dependences of $\rho_{ab}$ for at $T = 2.1$ K and various magnetic field in the range 0.50–2.0 T. At 0.9 T, $\phi$ dependences of $\rho_{ab}$ shows uplifts around $\phi = 45°$ and 225°, indicating two-fold symmetry of the in-plane anisotropy of $\rho_{ab}$. At 1.0 T, clear two-fold symmetry of $\rho_{ab}$ was observed. With increasing magnetic field, superconductivity was suppressed and two-fold symmetry of $\rho_{ab}$ also disappeared. Therefore, the NSC-like feature does not depend on the magnitude of magnetic field.

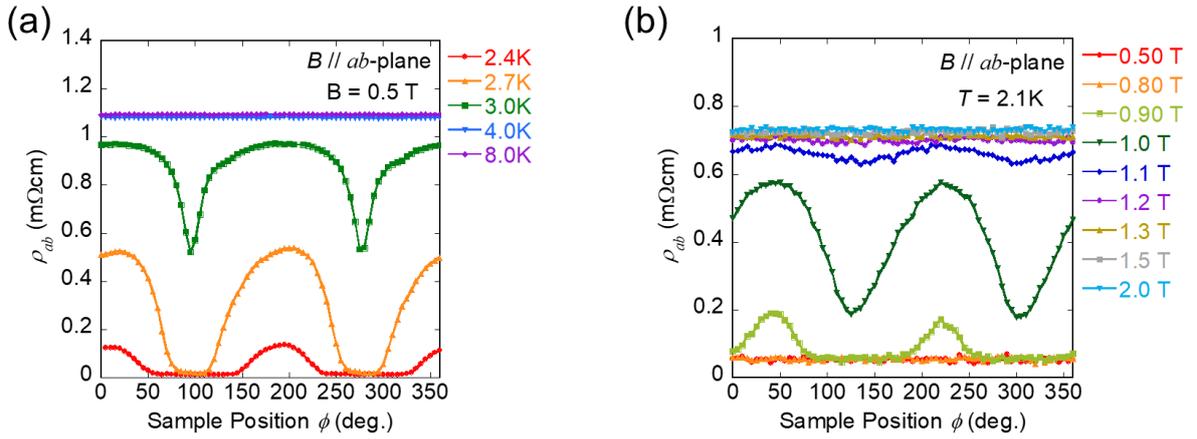

**Figure 6**: (a) $\phi$ dependences of the *ab*-plane resistivity ($\rho_{ab}$) for a CeOBiS$_{1.7}$Se$_{0.3}$ single crystal at $B = 0.5$ T and various temperatures in the range 2.4–8.0 K. (b) $\phi$ dependences of the *ab*-plane resistivity ($\rho_{ab}$) for a CeOBiS$_{1.7}$Se$_{0.3}$ single crystal at $T = 2.1$ K and at various magnetic fields ranging 0.50–2.0 T.

As introduced in the introduction, *in-plane chemical pressure* (CP) is a factor possibly essential for the emergence/disappearance (switching) of NSC in BiCh$_2$-based systems. This factor suppresses the weak local structural disorder peculiar to BiCh$_2$-based systems. Therefore, it is motivated to investigate in-plane local structure and in-plane anisotropy in another BiCh$_2$- based systems. The discovery of new NSC phase of BiCh$_2$-based superconductor shown in this study will be useful to address the issue.

## 4. Conclusion

We have synthesized CeOBiS$_{1.7}$Se$_{0.3}$ single crystals using a flux method. The single crystal X-ray structural analysis revealed that the crystal has a tetragonal structure with the space group of *P*4/*nmm*. BVS calculations showed that the valence of Ce ions is 3.25, indicating the mixed-valence state of Ce. Bulk superconductivity with a transition temperature of 3.3 K was confirmed through the electrical resistivity and magnetization measurements. The estimation of anisotropy parameter $\gamma$ of $B_{c2}$ revealed that the $\gamma$ for CeOBiS$_{1.7}$Se$_{0.3}$ is lower than that observed for RE(O,F)BiS$_2$ crystals. In the in-plane anisotropy measurements, two-fold symmetric in-plane anisotropy of magnetoresistance in the superconducting states was observed. These results suggest that CeOBiS$_{1.7}$Se$_{0.3}$ is a new nematic superconductor.

**Acknowledgements**
The authors thank A. Yamashita, T. D. Matsuda and O. Miura for supports in experiments. This work was partly supported by Tokyo Metropolitan Government Advanced Research (Grant Number: H31-1) and JSPS-KAKENHI (Grant Numbers: 19K15291 and 18KK0076).